# Multipurpose High Frequency Electron Spin Resonance Spectrometer for Condensed Matter Research

Kálmán Nagy, Dario Quintavalle, Titusz Fehér, András Jánossy

*(Institute of Physics and Condensed Matter Physics Research Group of the Hungarian Academy of Sciences, Budapest University of Technology and Economics)*

We describe a quasi-optical multifrequency ESR spectrometer operating in the 75–225 GHz range and optimized at 210 GHz for general use in condensed matter physics, chemistry and biology. The quasi-optical bridge detects the change of mm wave polarization at the ESR. A controllable reference arm maintains a mm wave bias at the detector. The attained sensitivity of $2 \times 10^{10}$ spin/G/(Hz)$^{1/2}$, measured on a dilute Mn:MgO sample in a non-resonant probe head at 222.4 GHz and 300 K, is comparable to commercial high sensitive X band spectrometers. The spectrometer has a Fabry-Perot resonator based probe head to measure aqueous solutions, and a probe head to measure magnetic field angular dependence of single crystals. The spectrometer is robust and easy to use and may be operated by undergraduate students. Its performance is demonstrated by examples from various fields of condensed matter physics.

# 1. Introduction

Once the tool of a few specialists, ESR at high frequencies is becoming a widely used method for research in chemistry, biology and physics. There are several reasons for the popularity of high frequency spectrometers, the most important one is higher spectral resolution. In many cases resonance shifts are proportional to the frequency while line widths are frequency independent and there is an improvement in the resolution of otherwise overlapping lines. In other cases the magnetic field or frequency dependent properties of the materials are of interest. E.g. in ferro- and antiferromagnets the frequency versus resonant magnetic field mode diagram and the frequency dependence of the magnetic relaxation are studied to understand the spin dynamics. In spin labeled biological systems the "time window" of the observed molecular motion is changed by varying the excitation frequency.

In the last decade a number of major scientific laboratories have developed high field ESR instrumentation. Here we quote only some of them to illustrate the diversity of applications and methods. Spectrometers for biological research (Cornell [1,2,3], Berlin [4, 5], Frankfurt [6,7]), and for doped semiconductors (Leiden [8]), have high sensitivities, while others are for general solid state physics like spectrometers operating at EPFL Lausanne [9], IFW Dresden, St. Andrews [10], Tallahassee, Pisa and Grenoble [11]. There are laboratories using pulsed magnets with typical fields up to 70 T, e.g. in Japan [12] and in Europe (Dresden-Rossendorf [13], Toulouse [14]).

Since the magnetization of samples increases with magnetic field and detection sensitivity is also better at higher frequencies, mm-wave ESR spectroscopy is in principle more sensitive than at the traditional cm wave length. In practice the comparison of sensitivities at different frequencies depends on several factors, e.g. a powder with g-factor anisotropy has a broader and thus smaller amplitude line at high frequencies. As we show in this paper, for many systems of interest a similar sensitivity may be obtained with rather simple means without resonant cavities at high frequencies like at X band using cavities. In sophisticated spectrometers with well designed mm-wave resonant cavities and carefully balanced quasi-optical bridges, the sensitivity can be much higher at high frequencies. High frequency resonant cavities necessary for the ultimate sensitivity may be, however impractical as they are difficult to construct and cumbersome to operate.

The spectrometer described here has been operated by many users, often undergraduate students for whom this was the first encounter with sophisticated experimental research. The idea is to have an instrument with high sensitivity, good resolution and at the same time robust enough to support the usual mistakes of beginners. It turns out that it is not more difficult to use a spectrometer at high frequencies than commercial X band spectrometers except for the need for handling cryogenic liquids to cool the superconducting magnet and the detector and to vary the temperature of the sample.

## 2. High frequency spectrometer setup

A block diagram of the spectrometer is shown in Fig. 1. The operation is based on the "induction" or "polarization coding" method where the change in polarization of the mm waves reflected from the sample is measured as the magnetic field is swept through the resonance. The linear polarization of the wave is carefully maintained from the source to the sample. The reflected perpendicularly polarized wave generated at resonance is separated from the incoming wave and detected. The outgoing wave is well isolated from the incoming wave, ideally off resonance no mm-wave power arrives from the sample to the detector; in our spectrometer the isolation is about 20 dB. For optimal operation the detector is biased by a phase and amplitude controlled reference arm. The magnetic field or the exciting mm wave amplitude is modulated at audio frequencies and the mm wave homodyne detection is followed by a lock-in amplifier. A PC controls data collection and measurement.

First we describe the magnet in more detail. Next we present the microwave system in a logical order following the light path. Then we discuss the probe heads and lastly the vibration isolation and the supporting structure.

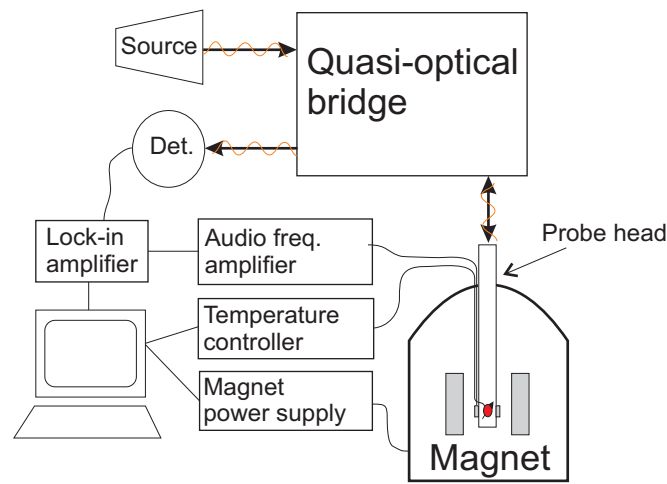

Fig. 1. Block diagram of the spectrometer. The quasi-optical bridge transmits the microwaves from the source to the probe head. It filters the signal from the reflected radiation and directs it to the detector. We use lock-in detection: the audio frequency amplifier drives a small coil in the probe head that modulates the magnetic field and the signal of the detector is processed by a lock-in amplifier. The measurement is controlled by a PC.

*2.1 The magnet*

The spectrometer is built around a 9 tesla superconducting magnet (Oxford Instruments) with a homogeneity specified as $10^{-5}$ in a volume of 1 cm$^3$ at the center of the magnet. This allows to resolve narrow lines, e.g. 0.03 mT broad lines were detected in 5.4 T magnetic field in few mm large samples of dilute $C_{59}N$ in solid $C_{60}$ [15]. The stability of the Oxford Instrument PS120[10] power supply is insufficient for such high spectral resolution and in this type of measurements the magnet is switched to persistent mode and the magnetic field is swept by the modulation coil in the probe head. A home built power supply provides simultaneously the audio frequency modulation and the magnetic field sweep of up to a few tens of mT.

The magnet is fitted with a variable temperature insert (VTI) supplied with liquid helium from the helium bath of the magnet. A stainless steel vessel in the VTI with a copper cap at the lower end ensuring good thermal coupling separates the sample space from the helium gas (or liquid) for temperature regulation. This vessel reduces pressure fluctuations in the sample space and thus increases the stability of the mm wave amplitude and phase. Also, it eliminates the risk of blocking the needle valve with air or ice during sample change, a time consuming and often costly failure.

*2.2 Microwave sources*

The spectrometer has two microwave sources and a set of frequency multipliers. The available frequencies and the corresponding resonance fields for *g*=2 are: 75 GHz (2.7 T), 111.2 GHz (4.0T), 150 GHz (5.4 T), 222.4 GHz (7.9 T) and 225 GHz (8 T). One source (Virginia Diodes) is based on a high stability phase locked dielectric resonator oscillator (DRO) at 13.9 GHz followed by one active doubler and a cascade of passive doublers. It emits 46 mW at 222.4 GHz and has a 111.2 GHz stage at much higher power. The second source (Radiometer Physics) is based on a 75GHz phase locked Gunn oscillator and has a frequency doubler and a tripler. Power is approximately 30 mW at 75 GHz and 1 mW at 225 GHz.

The oscillators are followed by a PIN diode for microwave amplitude modulation. This is very useful when matching the reference arm of the bridge to the operating point of the detector and for adjusting the excitation power level. Also, microwave amplitude modulation (chopping) with the PIN diode is used in some cases instead of field modulation. This technique is advantageous when the spectrum is much broader than the maximum field modulation amplitude, but it requires a very high mechanical stability of the spectrometer.

*2.3 Quasi-optical bridge*

The merits of reflection spectroscopy with polarization coding at mm wavelengths where resonant cavities are impractical were recognized early [16]. The main advantage is that the ESR signal is observed on a low power background independent from the excitation power. The recent development of very low phase noise powerful sources renders a well designed quasi-optical bridge indispensable. In a well balanced bridge, the background off-resonance power, reflected to the detector from the probe head or leaking through other paths, is small and a reference arm is used to bias the detector. The power levels at the sample and the detector are adjusted independently.

Our spectrometer operates in reflection mode with a quasi-optical bridge fabricated by Thomas Keating Ltd. The schema of the optical setup is presented in Fig.2. For simplicity the elliptical mirrors that periodically refocus the beam are omitted, and the magnet bore is rotated into the plane of the paper to avoid the overlapping of components. The source (1) emits a Gaussian beam linearly polarized at 45° with respect to the optical table. An isolator (2) that consists of a wire-grid polarizer, a 45° Faraday rotator and an absorber, eliminates standing waves and puts the polarization parallel to the table. A rotating wire-grid polarizer (3) splits the reference from the main beam. Right after the grid, the polarization in the reference arm is close to vertical while in the measuring arm it is almost horizontal. A moveable pair of plain mirrors (4) is used to change the length of the light path and thus the phase between the reference and the signal. An attenuator (5) in the measuring arm reduces the microwave intensity, e.g to avoid saturation of the ESR in samples with long spin relaxation times. A wire-grid polarizer (6) reflects the beam into the cylindrical corrugated waveguide of the probe head (7). At resonance the sample selectively interacts with one of the circularly polarized components of the linearly polarized microwave excitation and the reflected beam becomes elliptically polarized [17]. The wire-grid above the probe head (6) separates the resonance signal from the reflected wave. It transmits the component polarized orthogonally to the incident beam towards the detector, while it reflects the rest of the mm waves back to the isolator (2). The Faraday rotator (8) in the reference arm rotates the polarization from vertical to horizontal so that it can be joined with the signal beam by a wire-grid (9). A rotating wire-grid polarizer (10) adds the signal and the reference. The same polarizer (10) and a 45° Faraday rotator (11) form an isolator before the detector which eliminates standing waves caused by reflections from the detector (12).

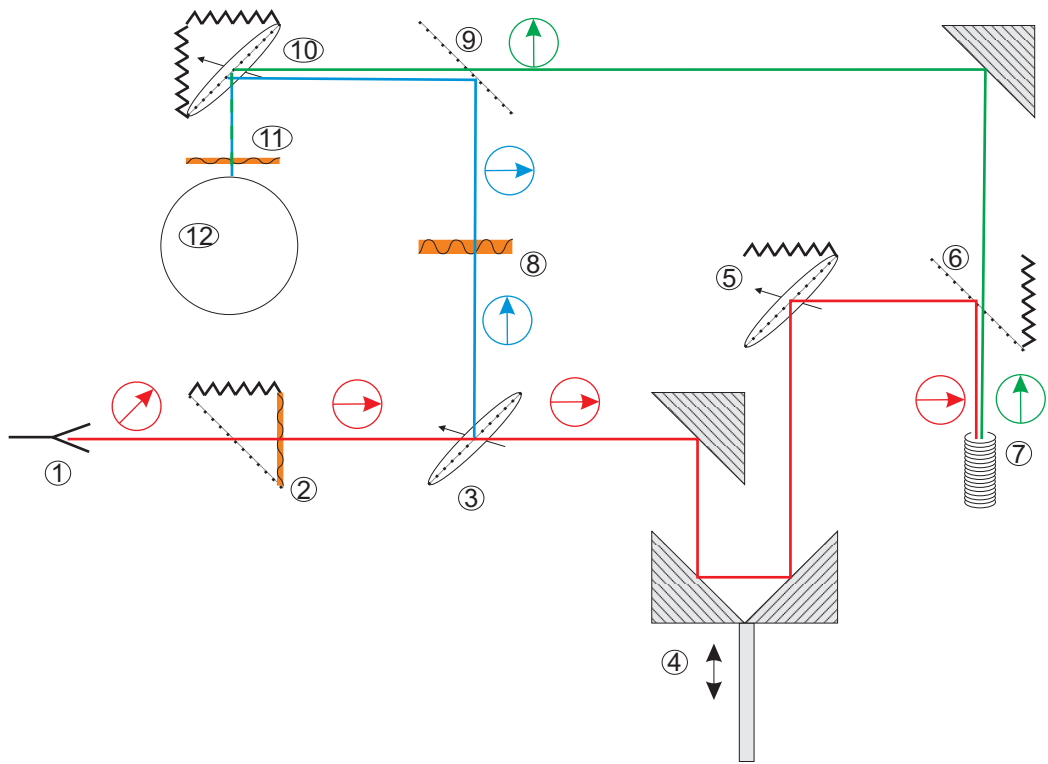

Fig. 2. Schema of the quasi-optical bridge (the focusing mirrors are omitted for clarity and the bore of the magnet is rotated into the plane of the drawing). The angle of the arrows in circles with respect to the horizontal direction corresponds to the approximate orientation of the polarization with respect the optical table. Dotted lines represent wire grid polarizers, the orientation of the ones marked with an arrow (i.e. 3, 5, 10) is adjustable. The zig-zag lines are dampers. The microwave radiation is emitted from the (1) microwave source, passes through the (2) isolator (45° Faraday rotator), (3) beam splitter, (4) phase shifter, (5) attenuator and enters the (7) probe head through the (6) grid for "polarization coding". The reference signal is directed towards the (8) 90° Faraday rotator, and then joins the signal with orthogonal polarization at grid (9). Finally the signal and the reference, added at the (10) rotating grid, enter the (12) detector isolated from the bridge by the (11) 45° Faraday rotator.

There are two detectors available, a hot-electron bolometer (QMC, QFI/2) for frequencies above 111.2 GHz and a Schottky diode (Millitech, DPX-12-RNFW0) for lower frequencies. The working point of the bolometer is at 3 mV signal (without preamplification), while that of the Schottky diode is at around 10 mV. The corrugated tube probe head and the isolators are designed for 210 GHz operating frequency. The spectrometer operates at lower frequencies with degrading efficiency and at 75 GHz the performance of the bridge is poor.

*2.4 Probe heads*

An oversized corrugated cylindrical waveguide transmits the mm waves from the bridge to the sample holder with very low losses keeping the linear polarization unchanged. This waveguide is connected to the sample holder block by a corrugated tapered section. The sample holder block and the interchangeable modulation coils are designed individually for different types of experiments and samples. The standard sample holders are 1.9 mm, 4.4 mm and 7 mm diameter cylinders for various frequencies and sample sizes with a movable mirror at the end to ensure that the sample is at the antinode of the microwave magnetic field (Fig. 3.). A 4.4 mm cylinder with a horizontal single-axis goniometer serves for the rotation of single crystals. Moving parts in the sample holder are manipulated by shafts from outside the cryostat. A low quality factor Fabry-Perot resonator has been developed for aqueous samples.

The temperature range of measurements is 2–340 K, but temperatures up to 520 K can be obtained in a special high temperature probe head. A high pressure mm-wave ESR probe head for a broad temperature range has been described [18] for a similar spectrometer at the Ecole Polytechnique Federale de Lausanne. Applying high pressure is simpler at high frequencies than in conventional X band spectrometers as the intrinsic sensitivity is higher and shorter wavelengths are more easily coupled to the generally constricted pressure cells.

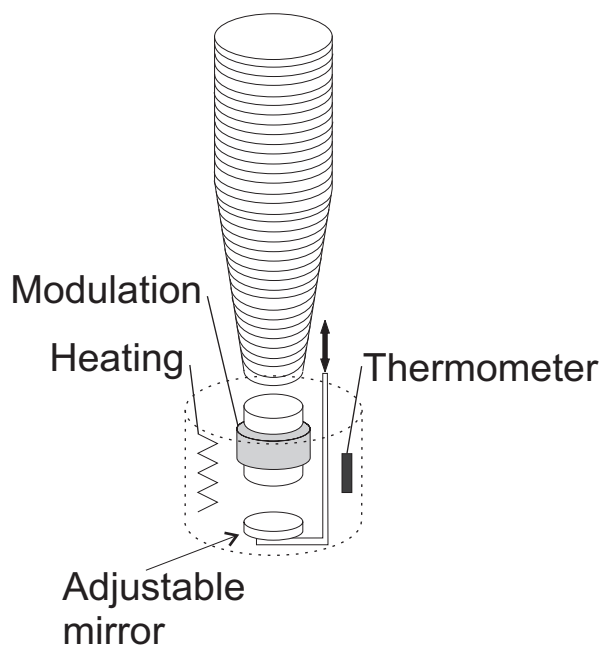

Fig. 3. Schema of the probe head. The corrugated waveguide is tapered near the end and connects to a copper cylinder; a moveable mirror inside the cylinder terminates the light path. The modulation coil is wound around the copper cylinder and the whole assembly is mounted inside a copper block (dashed cylinder) also housing the thermometer and the heater.

*2.5 Mechanical stability*

We made a great effort to reduce mechanical vibrations. The entire setup, magnet and bridge, is placed on a 2000 kg concrete platform lying on elastic copolymer blocks (CMD). Mechanical vibrations are damped efficiently above the self-resonance of the structure at about 10 Hz. Four brick pillars raise the optical bridge above the magnet. The optical bridge is mounted on top of the pillars on an aluminum support (Item Industrietechnik GmbH) that permits levelling and fine adjustment of the bridge position in three orthogonal directions with respect to the probe head. To change the sample, the bridge is retracted from the magnet on a rail. Stages for the operators built around the spectrometer are mechanically isolated from the spectrometer. To avoid degradation of the magnetic field homogeneity, no ferromagnetic material was used in the supporting structure.

## 3. High frequency ESR studies demonstrating spectrometer performance

The following few examples demonstrate the performance of the spectrometer and the potential of high frequency ESR spectrometry for solid state physics.

*3.1 Spectrometer sensitivity*

The sensitivity of the spectrometer was tested on Mn:MgO at ambient temperature. The concentration of $Mn^{2+}$ ions in the Mn:MgO powder, 1.58 ppm, was determined by comparing the ESR intensity at X band with a known amount of $CuSO_4 \cdot 5H_2O$ reference. The absolute sensitivity of the spectrometer at 222.4 GHz, calculated for spin 1/2 probes at 300 K is $2 \cdot 10^{10}$ spin/G/(Hz)$^{1/2}$. This is comparable to the sensitivity of modern, commercial X-band spectrometers (Fig. 4.). The corrugated waveguide and Faraday rotators are strongly frequency dependent and the sensitivity is lower at other frequencies; at 111.2 GHz it is about $4 \cdot 10^{11}$ spin/G/(Hz)$^{1/2}$. The mechanical stability of the spectrometer is excellent, the baseline is flat throughout the 0 to 9 tesla sweep.

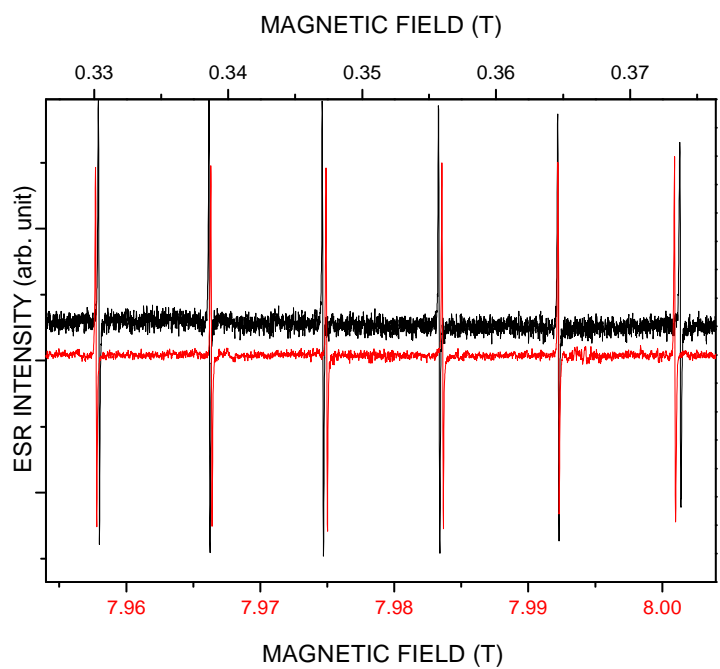

Fig. 4. Comparison of the room temperature spectra of 1mg Mn:MgO powder (Mn concentration 1.58 ppm) recorded in a commercial Bruker Elexsys E500 at 9.4 GHz (upper curve) and in the present spectrometer at 222.4 GHz (lower curve) showing comparable sensitivities. Right panel: the line widths are the same, 0.1 mT, at both frequencies, instrumental broadening does not deteriorate the large high frequency resolution.

*3.2 Phase segregation in $Na_2C_{60}$*

The ESR study of the intercalated fulleride salt with nominal composition $Na_2C_{60}$ but in fact inhomogeneous on a small length scale, demonstrates the importance of high spectral resolution. This fulleride was intensively studied as a possible realization of a Jahn–Teller–Mott insulator ground state with a continuous transition to a metal above ambient temperatures. The contradictory results reported in the literature [19,20], motivated the reinvestigation of the material.

The HF-ESR measurements combined with other experimental techniques [21] showed that the unusual behavior of $Na_2C_{60}$ at low temperature is a consequence of a phase segregation into phases with different Na concentrations. The phase segregation is driven by the diffusion of the $Na^+$ ions above about 450 K. The typical 3-10 nanometer domain size of the segregated phases is below the spatial resolution of X ray measurements. ESR, however, resolves three ESR active phases as their g factors are slightly different and at high frequencies the spectrum is split. As shown in Fig. 5, in the 225 GHz ESR the lines of the three phases (Phase 1, 2 and 3) are spread over 10 mT. In contrast, at 9 GHz there is a single structureless line, as the spectral resolution is 25 times less than at 225 GHz and the g-factor splitting is less than the ESR line width of individual phases.

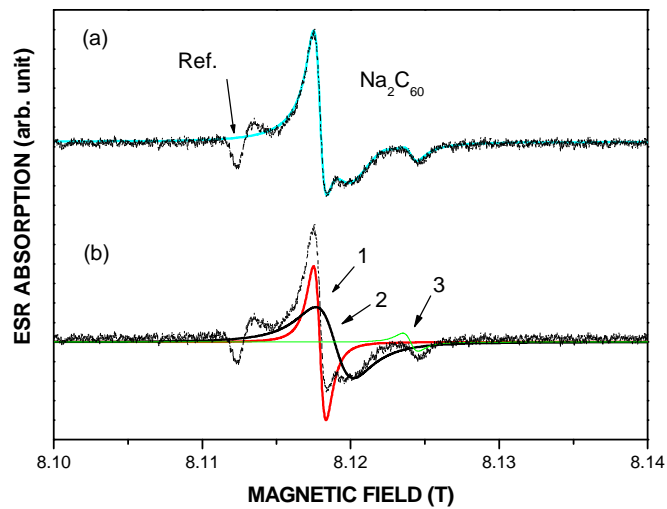

Fig. 5. 225 GHz ESR spectrum of the phase segregated alkali fulleride salt with nominal composition $Na_2C_{60}$, measured at 225 K. (a) The complex spectrum indicates phase segregation. The continuous line is a fit of the sum of three phases to the spectrum. (b) Decomposition of the spectrum into absorption lines of the three ESR active phases [21].

The temperature dependence of the susceptibilities is different for each segregated phase. The spin susceptibility is proportional to the absorption ESR line intensity, which is difficult to measure at high frequencies. Therefore ESR and SQUID measurements on the same sample between 4 and 450 K were combined. We determined the relative intensities of the ESR components and scaled the total ESR intensity to the static susceptibility data measured by SQUID. With this method, we determined the temperature dependence of the susceptibility in the various phases. The static susceptibility of Phase 1 has a Curie-like temperature dependence, the unusual temperature dependence of Phase 2 suggests that it is a mixture of phases with unresolved ESR lines even at 225 GHz. The approximately temperature independent susceptibility of Phase 3 shows that it is a metal.

*3.3 Single ion anisotropy in a magnetically dense anisotropic crystal*

ESR is a common method to measure the "crystal" or "zero field parameters" that characterize the magnetic anisotropy and the charge distribution surrounding magnetic ions. In dilute systems these parameters are given by the fine structure splitting of the ESR of the magnetic ion. In magnetically dense systems the exchange interaction between ions narrows the fine structure and only a single ESR line appears. The weak crystal fields of e.g. half filled electron shell ions do not affect the low frequency ESR in magnetically dense systems. At high magnetic fields and low temperatures, however, the most important crystal field parameters can be determined from the high frequency ESR. In this case the Zeeman splitting is large compared to the thermal energy, $k_B T$, and the ESR is at the crystal field shifted transition between the lowest energy Zeeman levels.

This method has been used [22] to determine the Mn crystal field parameters in the magnetic radical cation salt $ET_2MnCu[N(CN)_2]_4$ [23], where ET stands for bis(ethylenedithio)tetrathiafulvalene. The three dimensional polymeric anion network in this salt is unique among the $ET_2X$ materials. The magnetism arises mainly from two-dimensional layers of 5/2 spin $Mn^{2+}$ ions. The 1/2 spin ET cations are weakly coupled to the $Mn^{2+}$ ions and play little role in the low temperature ESR.
A temperature dependent anisotropic ESR shift was observed below 150 K [Fig. 6(a)]. As explained below, the shift is related to the distortion of the local environment of the $Mn^{2+}$ ions. In the spectrum calculated in the absence of exchange between the ions, the crystal field splits the $Mn^{2+}$ resonance into five allowed transitions between the six S=5/2 Zeeman levels [Fig. 6(b)]. The relative intensities of the five lines change with temperature as the population of the various Zeeman levels rearranges according to the Boltzmann distribution. At 300 K the spectrum is symmetric around the unshifted -½ –> +½ transition and the intensities of all 5 transitions are of the same order of magnitude. At low temperatures only the -5/2 –> -3/2 transition has significant intensity.

The exchange interaction between the $Mn^{2+}$ ions narrows the spectrum into a single line at the intensity weighted average of the 5 transitions. Hence the temperature dependence of the relative intensities of the fine structure lines transforms into a temperature dependent shift of the observed exchange narrowed line. A numerical analysis yielded the second order crystal field parameters. The exchange coupling between the $Mn^{2+}$ ions was also estimated from the line width.

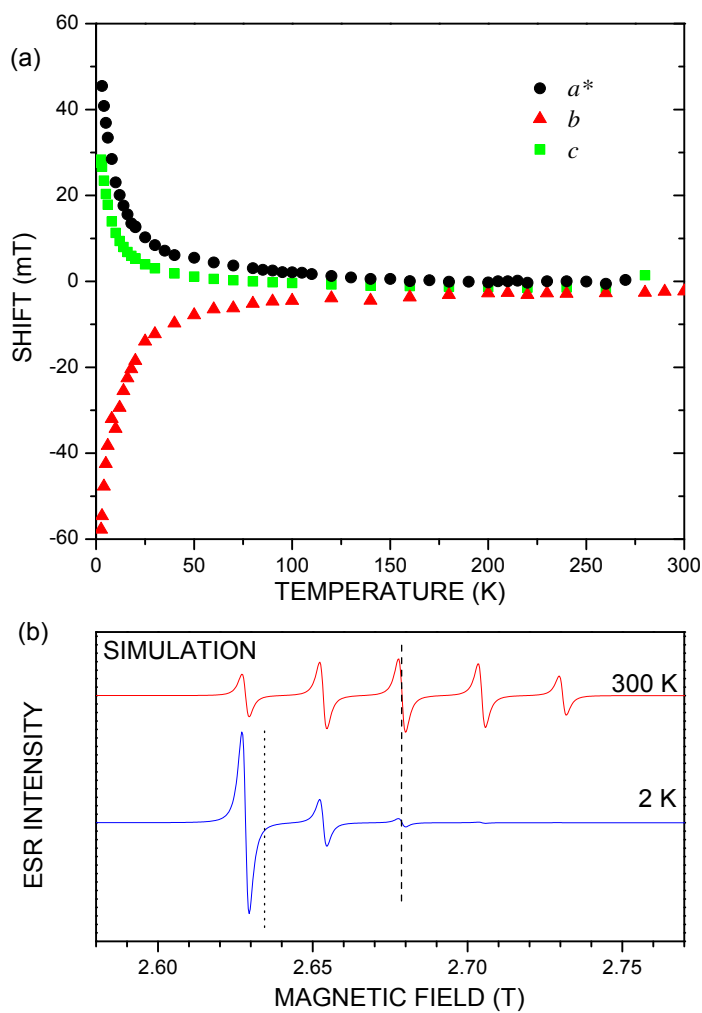

Fig. 6. Determination of the main crystal field parameters from the temperature dependent high frequency ESR shift of an ET$_2$MnCu[N(CN)$_2$]$_4$ single crystal. (a) Measured shifts in three orthogonal directions at 75 GHz. Anisotropic shifts at low temperature are attributed to an exchange-narrowed fine structure of the Mn$^{2+}$ ions. (b) Simulated Mn$^{2+}$ fine structure spectra at 300 K and 2 K. Dashed and dotted lines: the exchange coupling narrows the spectrum into a single line that is shifted by the crystal field at low temperatures [22].

*3.4 Antiferromagnetic mode diagrams*

High frequency magnetic resonance is a most powerful method to determine the interactions in magnetically ordered materials. In a system with more than one sublattice magnetizations (i.e. more than a simple ferromagnet) each sublattice magnetization is affected by the molecular fields of the other sublattices. As a consequence, the equilibrium magnetization orientations are complicated functions of the external magnetic field. Furthermore, when exciting the system by an oscillating magnetic field, the oscillating sublattice magnetizations produce extra oscillating fields of the same frequency at other sites, resulting in a coupled system with as many resonance modes as the number of sublattices.

Magnetic resonance in the layered organic charge transfer salt, κ-ET$_2$Cu[N(CN)$_2$]Cl illustrates the point. It has two structurally different but symmetry-related and chemically equivalent layers in the basic unit. Below the Neel temperature at 27 K, it is magnetically ordered. The four resonant modes found in a recent high frequency ESR study [24] confirmed the suggestion of Smith et al. [25] that it is a four-sublattice canted antiferromagnet with strong intra-layer interactions and several orders of magnitude smaller inter-layer interactions.

Finding the weak and narrow resonances in an antiferromagnet or a weak ferromagnet is not a simple task. In κ-ET$_2$Cu[N(CN)$_2$]Cl, isotropic exchange, Dzyaloshinskii-Moriya and anisotropic interactions shift the resonant fields by several teslas in a non-trivial way [Fig. 7(b)], and only the high sensitivity of the spectrometer and hints on the most suitable frequencies and magnetic field orientations given by model calculations made the observation [24] of all four magnetic eigenoscillation modes possible. The identification of the branches of the mode diagram requires measurements as a function of sample orientation. In Fig. 7(a) the dependence of two modes on the orientation of the magnetic field at 222.4 GHz is mapped. The rotation map of all four modes has been determined at 111.2 GHz. The data enabled to model with high precision the magnetic interactions between the four sublattices.

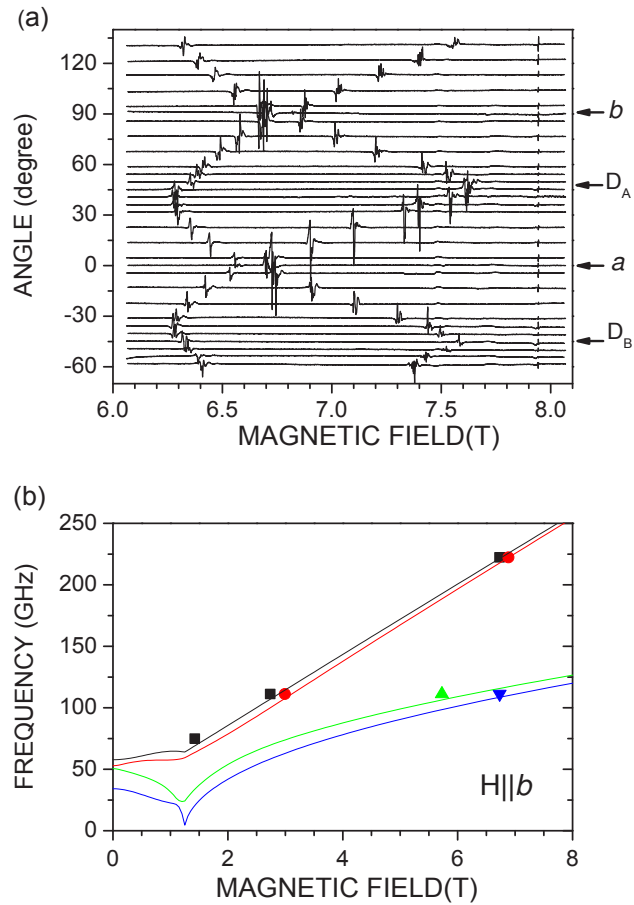

Fig. 7. Magnetic resonance in the four-sublattice canted antiferromagnetic insulator, κ-ET$_2$Cu[N(CN)$_2$]Cl at 4 K. (a) Spectra at 222.4 GHz as a function of magnetic field orientation in the (*a, b*) plane. The baseline of the spectra is at the angle between H and *a*. D$_A$ and D$_B$ are the Dzyaloshinskii-Moriya vectors of adjacent layers. (b) Resonance-field–frequency mode diagram for H ∥ *b*. Symbols represent observed resonances, lines are model calculations. There is only a limited frequency range where all 4 resonance modes of the 4 magnetic sublattices are observable [24].

*3.5 Fabry-Perot resonator*

ESR in water solutions is of scientific interest for structural biology but also challenging because of the strong absorption of microwaves in water. We developed a probe head with a Fabry-Perot resonator to measure water solutions at room temperature. The semi-confocal resonator has a fix coupling and the sample is placed on the flat mirror at the bottom of the resonator. The liquid sample is sealed by a screw-cap with a mylar window and a gasket made of teflon tape. The sample has a diameter of 4 mm and a thickness of approximately 30 μm. The calibration measurement shown in Fig.8 was done on a 10 mM solution of the stable free radical TEMPO (2,2,6,6-Tetramethylpiperidine-1-oxyl) in water at 222.4 GHz and room temperature. The signal to noise ratio was about 200 and the sensitivity of the spectrometer calculated from this spectrum is $7 \cdot 10^{11}$ spin/G/(Hz)$^{1/2}$.

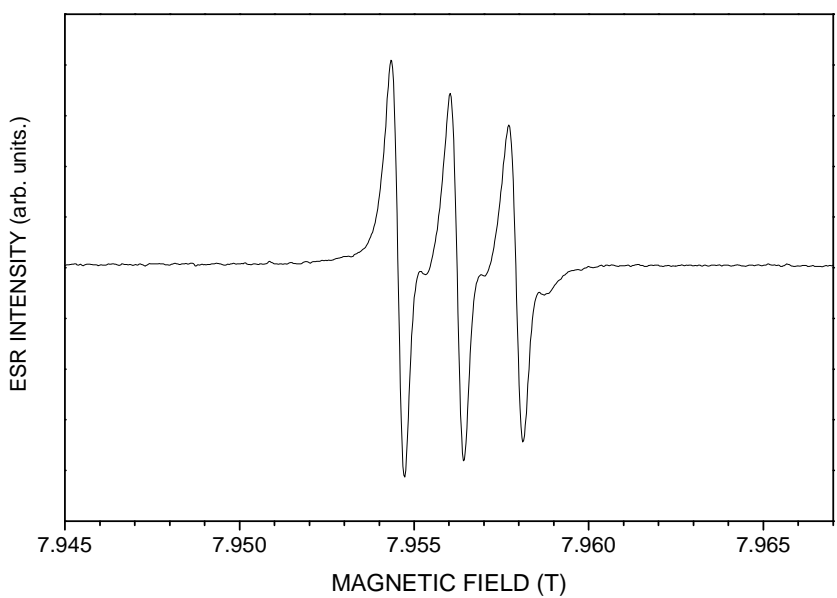

Fig. 8. Room temperature ESR spectrum of 10 mM water solution of the stable free radical TEMPO. The 222.4 GHz spectrum was recorded in a Fabry-Perot resonator.

## 4. Conclusions

High frequency ESR spectroscopy is an efficient research tool for condensed matter physics, chemistry and biology. Development of high-power and stable microwave sources and the availability of commercial quasi optical elements in the past decade brought it within the reach of a wide range of laboratories. The spectrometer described in this paper has a high spectral resolution, a high sensitivity, is simple to operate and is not very costly when compared to low frequency commercial spectrometers. The polarization coding with a controllable reference arm makes the operation at high frequencies similar to low frequency ESR bridges. Several frequencies are available and at the optimum frequency of 222.4 GHz the sensitivity is comparable to the best commercial X band spectrometers. This sensitivity is reached for low loss samples without resonant cavities, which are difficult to use at mm waves and limit sample size. The spectrometer works in a broad temperature range.

Adaptation of high pressure cells is simpler at mm waves than for standard X-band spectrometers.

The examples of ESR in fullerides, organic magnets and weak antiferromagnets illustrate the use of the spectrometer for various problems, where the higher resolution or higher magnetic fields are necessary to obtain the required information. Microwave losses are larger at mm waves in water than at lower frequencies and resonant structures are required for dilute aqueous solutions of spin labelled molecules.

**Acknowledgement**

We are indebt to Richard Wylde for his help in the design of the quasi-optical bridge. This work was supported by the Hungarian National Research Fund OTKA NK-60984, PF-63954, K-68807, NN-76727. T. F. acknowledges financial support from the János Bolyai program of the Hungarian Academy of Sciences.